\documentclass[conference]{IEEEtran}
\usepackage{cite}
\usepackage{amsmath,amssymb,amsfonts}
\usepackage{algorithmic}
\usepackage{graphicx}
\usepackage{textcomp}
\usepackage{xcolor}
\usepackage{subcaption}
\usepackage{booktabs}
\usepackage{bm}
\usepackage{multirow}

\ifCLASSINFOpdf
\else
\fi
\begin{document}
%
\title{Benchmarking the Performance of Large Language Models on the Cerebras Wafer Scale Engine}

\author{
\IEEEauthorblockN{ Zuoning Zhang\IEEEauthorrefmark{1}, Dhruv Parikh\IEEEauthorrefmark{1}, Youning Zhang\IEEEauthorrefmark{2}, Viktor Prasanna\IEEEauthorrefmark{1}}
\IEEEauthorblockA{
    \IEEEauthorrefmark{1}University of Southern California \\
    \IEEEauthorrefmark{1}\{zuoningz, dhruvash, prasanna\}@usc.edu \IEEEauthorrefmark{2}\{youningzhang\}@berkeley.edu}
}
\maketitle

\begin{abstract}
Transformer based Large Language Models (LLMs) have recently reached state of the art performance in Natural Language Processing (NLP) and  Computer Vision (CV) domains. LLMs use the Multi-Headed Self-Attention (MHSA) mechanism to capture long-range global attention relationships among input words or image patches, drastically improving its performance over prior deep learning approaches.  


\indent In this paper, we evaluate the performance of LLMs on the Cerebras Wafer Scale Engine (WSE). Cerebras WSE is a high performance computing system with 2.6 trillion transistors, 850,000 cores and 40 GB on-chip memory. Cerebras WSE's Sparse Linear Algebra Compute (SLAC) cores eliminate multiply-by-zero operations and its 40 GB of on-chip memory is uniformly distributed among SLAC cores, enabling fast local access to model parameters. Moreover, Cerebras software configures routing between cores at runtime, optimizing communication overhead among cores. As LLMs are becoming more ubiquitous used, new hardware architectures are required to accelerate LLM training and inference. We benchmark the effectiveness of this hardware architecture at accelerating LLM training and inference. Additionally, we analyze if Cerebras WSE can scale the memory-wall associated with traditionally memory-bound compute tasks using its 20 PB/s high bandwidth memory. Furthermore, we examine the performance scalability of Cerebras WSE through a roofline model. By plotting training throughput against computational intensity, we aim to assess their effectiveness at handling high compute-intensive LLM training and inference tasks.




\end{abstract}



%

\section{Introduction}

Large Language Models (LLMs) have been gaining growing interest and led to significant breakthroughs in natural language processing, enabling applications from automated text generation \cite{vaswani2017attention} and machine translation \cite{mccann2017learned} to conversational AI \cite{xue2022recent}. Their ability to understand human inquiry and generate human-like text has enabled advancements in fields such as education \cite{zhang2024simulating} and healthcare \cite{gebreab2024llm}, fostering human and AI to have closer and more direct interactions. 

\indent Before transformer models, Recurrent Neural Networks (RNNs) \cite{elman1990finding} and Long Short-Term Memory (LSTMs) were the state-of-the-art models for sequential data processing tasks. RNNs operate on sequential data processing tasks by maintaining a hidden state that captures information from previous inputs and handles one input element at a time. This enables RNNs to be able to handle inputs of arbitrary length, therefore making it suitable for language-related tasks. RNNs have been deployed in applications such as machine translation \cite{chaudhary2018bilingual} and speech recognition \cite{miao2015eesen}. However, RNNs suffer from vanishing/exploding gradients \cite{salehinejad2017recent} and struggle at handling inputs that exhibit long-range dependencies among input words. LSTMs address these issues by introducing a gating mechanism, allowing important information to be preserved over long sequences. This makes LSTMs to be more effective at handling input sequences with long-term dependencies \cite{sak2014long}. Despite the success of RNNs and LSTMs at handling textual and sequential data, they struggle with parallelization. This limits RNNs and LSTMs from scaling to large model sizes without sacrificing efficiency, therefore making them struggle to handle complex inputs and inputs with extremely long-range relationships. These limitations led to the development of transformer models \cite{vaswani2017attention}, which utilize a Multi-Head Self-Attention (MHSA) mechanism to process input tokens simultaneously in order to find relationships among input tokens. The parallel nature of MHSA enables the model to efficiently capture relationships among very long input sequences. For example, in GPT-3 \cite{brown2020language}, the context window size is 2048 tokens, meaning the model will find relationships among the previous 2048 tokens. Moreover, the parallel nature of transformer models allows them to scale in size and excel at handling complex input sentences.

\indent Two cutting-edge transformer models that utilize the MHSA mechanism are BERT \cite{devlin2018bert} and GPT \cite{brown2020language} models. Unlike traditional models that process input tokens sequentially (left-to-right or right-to-left), BERT uses transformer encoders, which utilize a bidirectional approach to process input sequences, meaning BERT can calculate MHSA values of a token based on tokens both to the left and right. This enables BERT to have greater contextual understanding of each word and discover more relationships within the input sequence. GPT models utilize transformer decoder components to process texts in a left-to-right manner, where each token can only compute MHSA based on previous tokens in the sequence. This unidirectional nature allows GPT models to predict the next word in a sequence based on the context of all preceding words, making it suitable for autoregressive text generation tasks. 

\indent However, the vast scale of these transformer-based LLMs architectures requires significant training and inference costs, presenting challenges in computational and economic resources. In 2018, BERT had a model size of 110M parameters \cite{devlin2018bert}. In 2020, GPT had 175B parameters \cite{brown2020language}. The model sizes increased by more than 1000 times in two years. The growing demand for training larger LLMs requires growing compute intensities and hardware resources. To address these challenges, several new and innovative hardware architectures have been proposed, aiming to optimize the efficiency of LLMs training and inference \cite{luk2024hardware} \cite{huang2024new}.

\indent One hardware architecture that aims to accelerate the LLMs training and inference tasks is the Cerebras CS-2 system. The Cerebras CS-2 system integrates an entire wafer-scale chip, Cerebras WSE-2, to build a powerful AI accelerator. This enables Cerebras WSE-2 to possess extremely high memory bandwidth and compute intensity. The WSE-2 chip is more than 46000  \( mm^2 \) in size with 2.6 trillion transistors. The Cerebras WSE-2 architecture is composed of 850,000 cores arranged in a 2-D mesh topology, enabling 20PB/s memory bandwidth and 220Pb/s fabric bandwidth \cite{lie2023cerebras} \cite{lauterbach2021path}.

\indent Analyzing the training throughput and inference latency of LLMs in the Cerebras WSE becomes a critical task as this architecture, with extremely high memory bandwidth, has the potential to allow LLMs training and inference to take advantage of its abundant compute resources without being bottlenecked by memory bandwidth. In this work, we make the following contributions: 

\begin{itemize}
    \item  \textbf{Training analysis:} We performed in-depth analysis of the training throughput of BERT, on classification task, and GPT-3, on autoregressive text generation task, across different model sizes and batch sizes on the Cerebras WSE platform. 
    \item \textbf{Inference analysis:} We perform the end-to-end inference latency analysis for the BERT model on binary classification task on the Cerebras WSE platform across different BERT model sizes and commonly used batch sizes.
    \item \textbf{Projected training epoch duration:} Based on our observed training throughputs, we performed analysis to obtain the projected per-epoch training time to train GPT-3 models and BERT models on the PILE dataset \cite{gao2020pile} and SST-2 dataset \cite{socher2013recursive}. 
    \item \textbf{Roofline model:} We completed roofline model analysis for training throughput of BERT and GPT-3 models across different model sizes to gain insight into the scalability of the Cerebras WSE on LLMs training. 
\end{itemize}

Our aim is to deepen our understanding of Cerebras WSE platform's potential to perform intensive LLMs training and inference tasks. 

Section \ref{sec:background} briefly introduces LLMs and their applications in different fields. Furthermore, this section briefly introduces current state-of-the-art hardware architecture, Cerebras WSE, for LLMs training and inference. Section \ref{sec:experiments} describes the models, model sizes, datasets used, and batch sizes for models being evaluated for the experiments, along with the computing platforms on which the experiments were performed. Results are analyzed in section \ref{sec:results}. Discussion and conclusion follow in section \ref{sec:conclusions}.

\section{Background}
\label{sec:background}


\subsection{Large Language Models (LLMs)}
\label{subsec:LMM}

Large language models have gained growing interest over the past few years. Especially with the release of ChatGPT \cite{openai2022chatgpt}, users began recognizing the capabilities of LLMs and leverage them to assist with daily tasks that were previously time-consuming. Because of its remarkable language understanding abilities, more than one million users signed up to use ChatGPT within one week of its release \cite{abdullah2022chatgpt}. Moreover, LLMs have created a profound influence in many industries and fields. For example, LLMs ability to generate and debug code can accelerate the software development process and enhance programmers' productivity \cite{gu2023llm}. LLMs can also assist with the creative writing process by offering new ideas and providing feedback and grammatical corrections on input writings of users \cite{abdullah2022chatgpt}. 

\indent Current LLMs use transformer architecture and utilize the Self-Attention (SA) mechanism to efficiently process and understand relationships among words in a sentence. This mechanism enables LLMs to have a powerful understanding of the meaning and context of a given sentence. Mathematically, SA can be defined as the following sequences of operations
\begin{equation}
\label{eq:self_attention_Q}
\bm{Q} = \bm{W}^{Q}\bm{X} 
\end{equation}

\begin{equation}
\label{eq:self_attention_K}
\bm{K} = \bm{W}^{K}\bm{X} 
\end{equation}

\begin{equation}
\label{eq:self_attention_V}
\bm{V} = \bm{W}^{V}\bm{X} 
\end{equation}

\begin{equation}
\label{eq:self_attention}
\bm{A} = \text{softmax}\left(\frac{\bm{Q}\bm{K}^{T}}{\sqrt{{d}_{k}}}\right) \bm{V}
\end{equation}

$ \bm{W}^{Q} \in \mathbb{R}^{d \times {d}_{k}}$, $ \bm{W}^{K} \in \mathbb{R}^{d \times {d}_{k}}$, $ \bm{W}^{V} \in \mathbb{R}^{d \times {d}_{v}}$ are the weight matrices for query, key, and value, respectively. 
$ \bm{Q} \in \mathbb{R}^{n \times {d}_{k}}$, $ \bm{K} \in \mathbb{R}^{n \times {d}_{k}}$, $ \bm{V} \in \mathbb{R}^{n \times {d}_{v}}$ are the query, key, and value matrices, respectively. 
$ \bm{X} \in \mathbb{R}^{n \times d}$ is the input sentence embedding matrix, where $n$ is the input sequence length, $d$ is the embedding length of each word in the input sentence. The SA output $\bm{A} \in \mathbb{R}^{n \times d_{v}}$ will then be projected to the same shape as input $\bm{X} $ through a feed-forward network. This sequence of operations define the essential elements of a transformer block and is also called a single attention head. In transformer-based LLMs, each transformer block can contain multiple such attention heads. Each head captures different relationships among the tokens in the input sequence, baking within the model a richer context and understanding via the Multi-Headed Self-Attention (MHSA) mechanism. As the input and output dimensions of a transformer block are the same, multiple transformer blocks can be stacked, allowing earlier transformer blocks to provide more context for later blocks. 

\indent Transformer models are pre-trained over a large corpus of tokens via tasks such as next word prediction (GPT) \cite{geva2022transformer} and masked language modeling (BERT) \cite{salazar2019masked}. 

\subsection{GPT}
\label{subsec:GPT}
GPT is a transformer decoder-only architecture where each input only computes SA among previous input tokens. Namely, when computing output SA token ${\bm{a}_{i} \in \mathbb{R}^{d_v}}$, SA only computes among ${\bm{q}_{i}} \in \mathbb{R}^{d_k}$ with ${\bm{k}_{j} \in \mathbb{R}^{d_k}}$, ${\bm{v}_{j} \in \mathbb{R}^{d_v}}$ where $j \leq i$. This ensures that the model only attends to past and current tokens, not future unseen ones. At each time stamp, GPT computes a probability distribution over the entire vocabulary to predict the probability of next word, enabling it to generate the next word. This process is also called auto-regressive decoding or text generation. Auto-regressive text generation during inference terminates when an \emph{endoftext} token is generated or a predefined maximum output sequence length is reached. 

\begin{figure}[htbp]
\centerline{\includegraphics[width = \linewidth]{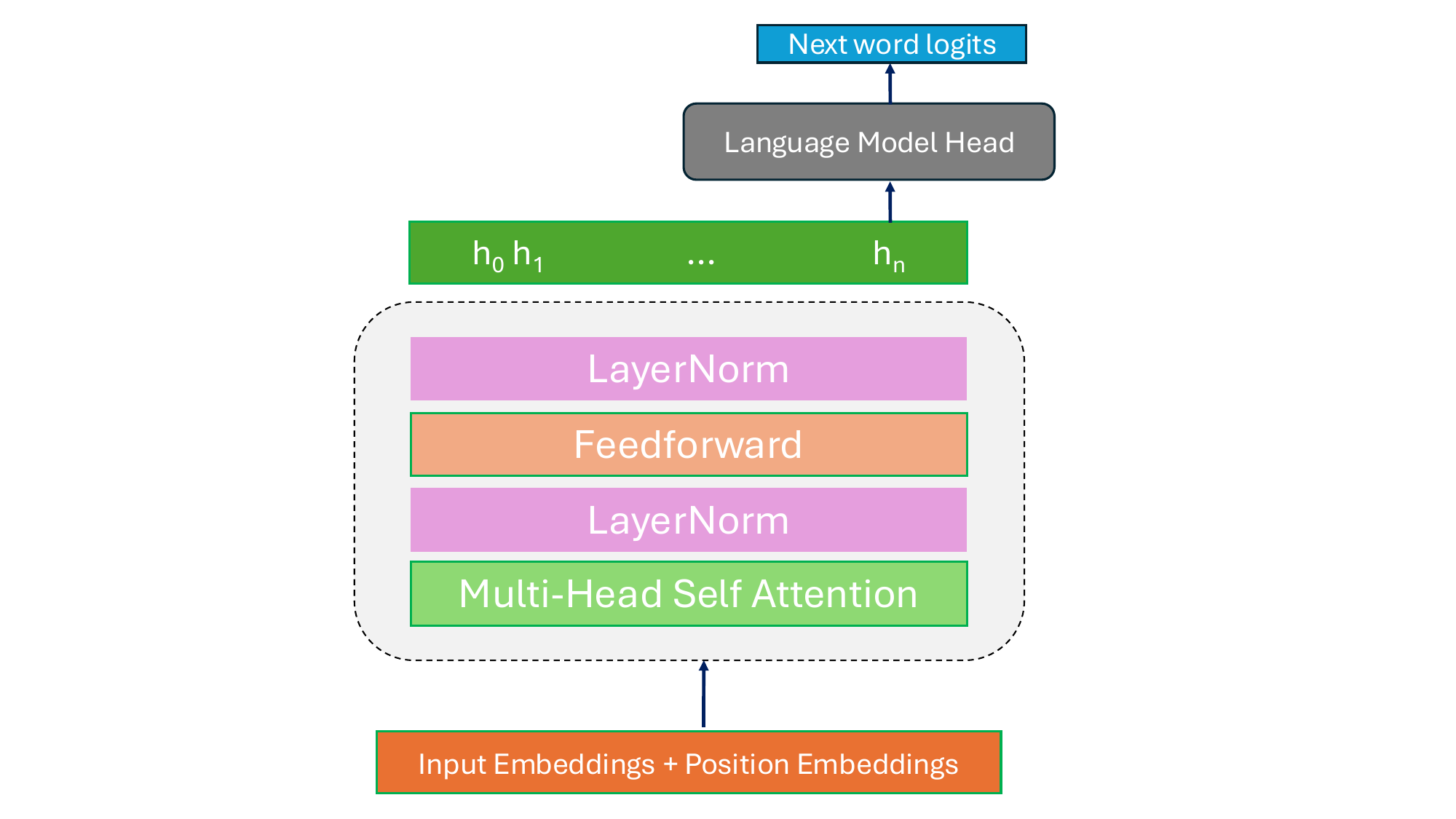}}
\caption{GPT transformer decoder architecture. Input embeddings are first added with positional embeddings and then fed into GPT transformer-decoder block, where MHSA is performed.}
\label{fig:gpt}
\end{figure}

\subsection{BERT}
\label{subsec:BERT}

BERT\cite{devlin2018bert} is a transformer encoder-only architecture where each input computes SA among both previous and future input tokens, called bidirectional SA.  Namely, when computing output SA token ${\bm{a}_{i} \in \mathbb{R}^{d_v}}$, SA only computes among ${\bm{q}_{i}} \in \mathbb{R}^{d_k}$ with ${\bm{k}_{j} \in \mathbb{R}^{d_k}}$, ${\bm{v}_{j} \in \mathbb{R}^{d_v}}$ where $j \leq n$, where $n$ is the input sequence length. Because SA is computed over the entire input sequence, BERT models have a global comprehensive context over all tokens in the input sequence. Normally, a [\emph{class}] token is added at the beginning of the input sequence. At the output layer, the final hidden state of the [\emph{class}] token is used for the classification of the entire sentence. This ensures that the final classification of the sentence is not biased toward any specific word in the input sequence. BERT models are typically pre-trained on tasks like Masked Language Modeling (MLM) \cite{devlin2018bert}, where a random subset of the input sequence is masked and the model predicts the masked words. BERT models are also pre-trained via Next Sentence Prediction (NSP), where the model predicts whether the given two sentences should follow each other. After pre-training, BERT models are fine-tuned for task-specific applications, such as sentiment analysis of input sequence \cite{sousa2019bert}. 

\begin{figure}[htbp]
\centerline{\includegraphics[width = \linewidth]{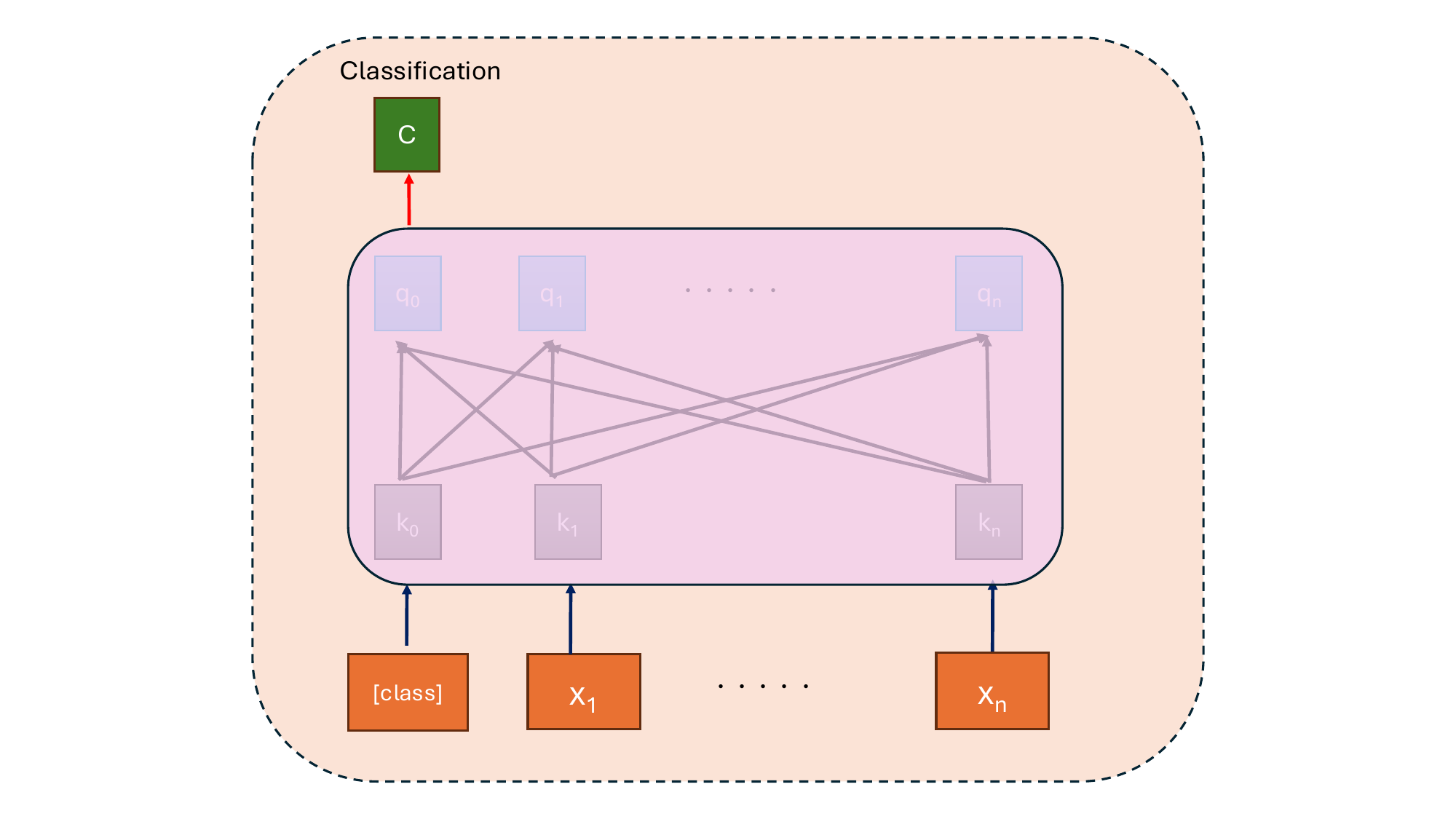}}
\caption{BERT transformer encoder architecture. [\emph{class}] token embedding is appended to the front of the input sequence. The BERT architecture is similar to the GPT model architecture shown in Figure \ref{fig:gpt}, except that MHSA occurs between the current token and tokens to the left and right. In BERT, the final output hidden state of [\emph{class}] token is used to predict the probability distribution for classification of the input sequence.}
\label{fig:bert}
\end{figure}

\subsection{Cerebras WSE}
\label{subsec:cerebras}

All training and inference experiments are conducted on the Cerebras Wafer Scale Engine (WSE) and Cerebras CS-2 system. Cerebras WSE is a powerful AI accelerator built to train LLMs \cite{sengupta2023jais}. As of May 2023, Cerebras WSE is the largest chip ever built, at 46,000 mm$^2$ and containing 2.6 trillion transistors and 850,000 cores. Cerebras WSE's large die size enables greater compute power and enables more computations to occur on-chip \cite{lauterbach2021path}. This reduces off-chip communications and enables computation resources to be fully utilized. Cerebras chip is built with Taiwan Semiconductor Manufacturing Company (TSMC) 7 nm process and runs at 1.1 GHz frequency \cite{lie2023cerebras}. The WSE's physical design contributes 50\% silicon area to static random access memory (SRAM) and 50\% to computation logic in each core. Each core features a local 48 KB SRAM that enables the core to have fast access to local memory. In each core, memory is organized into eight 6 KB banks, each 32-bit wide. On top of the local 48 KB memory, each core also has a 256-byte, 64-bit wide, software-configured cache, enabling even faster data access for most frequently used data \cite{lie2023cerebras}. This uniformed distributed on-chip memory architecture enables low memory access latency for each core and very high aggregated memory throughput. 

\indent Cerebras WSE's fine-grained data flow scheduling,  enables cores to only perform computations on non-zero data, saving dynamic power of cores \cite{lie2023cerebras}. The combination of this and the high memory bandwidth enables efficient computations on data with unstructured sparsity \cite{thangarasa2024mediswiftefficientsparsepretrained}, which is well suited for neural network computations as model weights and input data can possess arbitrary levels of sparsity. Moreover, on top of fine-grained data flow scheduling, each core also has 8 micro-threads holding 8 independent tensor instructions. The scheduler chooses among these cores to execute in the core compute logic, guaranteeing that there is always useful work ready for the compute logic to execute.

\indent Cores in Cerebas WSE are arranged in a 2-D mesh topology. Each core has a router that has a 32-bit bidirectional port to four adjacent cores in north, south, east, and west direction, as well as one 32-bit port to the compute logic within the core. Data packets in Cerebras WSE are 32-bit long, 16-bit data and 16-bit control. Packets are communicated through static routing. Each router has 24 local static routes that can be configured, called colors. The static routing mechanism and 2D mesh topology enable high-speed data communication among cores \cite{lie2023cerebras}. 

\indent Cerebras WSE also uses weight streaming to enable training very large models. Model weights are stored in an external memory device DRAM and flash memory called MemoryX. MemoryX sends the weights of each layer to Cerebras WSE at compute time. After Cerebras WSE computes the output values using the streamed weights and the data in its cores, backpropagation occurs and the gradients of the layer are sent back to MemoryX to perform weight updates. Storing model weights externally enables extremely large model sizes as they do not consume on-chip memory \cite{lie2023cerebras}. 

\indent Cerbras WSE's architecture enables high memory bandwidth, rich compute resources, and very low overhead communication between cores. This architecture enables efficient training and inference of very large neural network models.

\begin{figure}[htbp]
\centerline{\includegraphics[width = \linewidth]{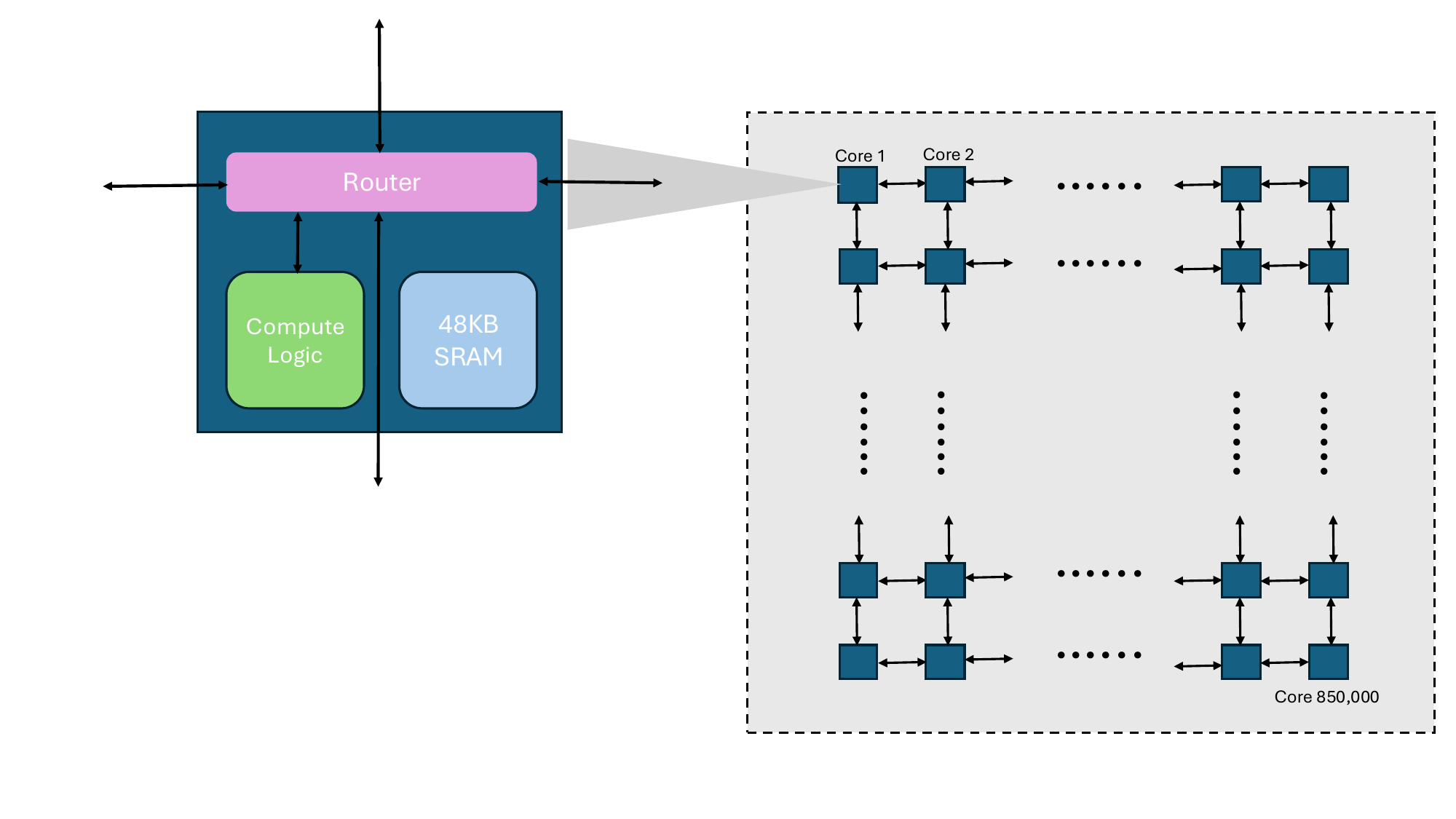}}
\caption{Cerebras WSE architecture. Cores are connected in 2D mesh topology. Each core has a dedicated router that connects to neighboring cores and its own compute logic. Each core also has 48 KB SRAM, totaling 40 GB on-chip SRAM on the entire chip }
\label{fig:cerebras}
\end{figure}

\section{Experiments}
\label{sec:experiments}




\subsection{Cerebras Platform Details}
\label{subsec:platform}
The training throughput and inference latency analysis for GPT-3 and BERT is conducted on the Cerebras CS-2 system. The CS-2 system contains the host CPU and the Cerebras WSE. Table \ref{tab:platform-specifications} contains details of these platforms.

\begin{table}[!ht]
\centering
\caption{Specifications of platforms }
\begin{tabular}{c|cccccc}
 \toprule
\textbf{Platforms} & Cerebras CS-2 & AMD EPYC 7702P   \\ 
\midrule \midrule 
 {Platform Technology}  & TSMC 7 nm   & TSMC 7 nm \\ 
{Frequency} & 1.10 GHz  & 2.0 GHz  \\ 
{Peak Performance}& 7.5 PFLOPS & 2.04 TFLOPS  \\ 
{On-chip Memory}& 40 GB & 256 MB L3 cache   \\
{Memory Bandwidth}& 20 PB/s & 204.8 GB/s   \\ \bottomrule
\end{tabular}
\label{tab:platform-specifications}
\end{table}





\subsection{Datasets}
\label{subsec:datasets}
We utilized the Stanford Sentiment Treebank (SST-2) dataset \cite{socher2013recursive} for training and inference of BERT models. We utilize the SST-2 dataset for fine-tuning of BERT model. The dataset is composed of movie reviews with binary classification tasks, classifying each movie review as positive or negative. We split the SST-2 dataset to have 67350 samples for training and 873 samples for evaluation, which is used to measure BERT inference latency. 

\indent We utilized the PILE \cite{gao2020pile} dataset for training of GPT-3 models. The PILE is an 825 GB dataset, containing 211,043,181 samples \cite{biderman2022datasheet} of English text dataset collected from 22 high-quality sub-datasets of different domains. By training LLMs on datasets of diverse domains, the model is able to gain greater general domain knowledge. We utilize the PILE dataset for pre-training of the GPT-3 models. The task completed by the model is to predict the next word based on all previous words in the sequence. Therefore, the label of each training sample is the input sequence tokens right shifted by one token. To simplify the training settings and reduce the time taken to complete the experiments, we use a 16 MB subset, containing 10,000 samples, of the PILE dataset for training of the GPT-3 models.




\subsection{Model Hyper-parameters}
\label{subsec:model-params}

We performed our experiments on the BERT-base (111M) and BERT-large (340M) variants. For GPT-3, we evaluated the 256M, 590M, 2.7B, 6.7B, 13B, and 20B model sizes. Table \ref{tab:bert-params} and Table \ref{tab:gpt-params} represent the hyper-parameters of each model variant. In these tables, $L$ represents the number of layers (transformer blocks), $D$ represents the hidden size, $H$ represents the number of self-attention heads and $MSL$ represents the maximum sequence length of the input. 

\begin{table}[!ht]
\centering
\caption{BERT hyper-parameters }
\begin{tabular}{c|cc}
 \toprule
\textbf{BERT } & Base & Large  \\ 
\midrule \midrule 
 {$L$}  & 12  & 24 \\ 
{$D$} & 768  & 1024 \\ 
{$H$}& 12 & 16  \\ 
{$MSL$}& 128 & 128   \\ \hline
\end{tabular}
\label{tab:bert-params} 
\end{table}

\begin{table}[!ht]
\centering
\caption{GPT-3 hyper-parameters }
\begin{tabular}{c|cccccc}
 \toprule
\textbf{GPT-3 } & 256M & 590M & 2.7B & 6.7B & 13B & 20B \\ 
\midrule \midrule 
 {$L$}  & 14  & 18 & 32 & 32 & 40 & 44\\ 
{$D$} & 1088  & 1536 & 2560 & 4096 & 5120 & 6144 \\ 
{$H$}& 17 & 12 & 32 & 32 & 40 & 64\\ 
{$MSL$}& 2048 & 2048  & 2048 & 2048 & 2048 & 2048\\ \hline
\end{tabular}
\label{tab:gpt-params}
\end{table}




\subsection{Performance metrics}
\label{subsec:performance_metrics}

For training of BERT and GPT-3 models, we use throughput, samples per second, as the performance metric. For inference of BERT models, we use end-to-end latency as the performance metric. Because BERT models perform classification tasks, we define the end-to-end latency of BERT models as the duration from when the input batch is given to the model to when the model output is produced.   
\section{Results}
\label{sec:results}

\subsection{BERT Training Performance Analysis}
\label{subsec:bert_train}

\begin{figure}[htbp]
\centerline{\hspace{-20pt}\includegraphics[width = \linewidth]{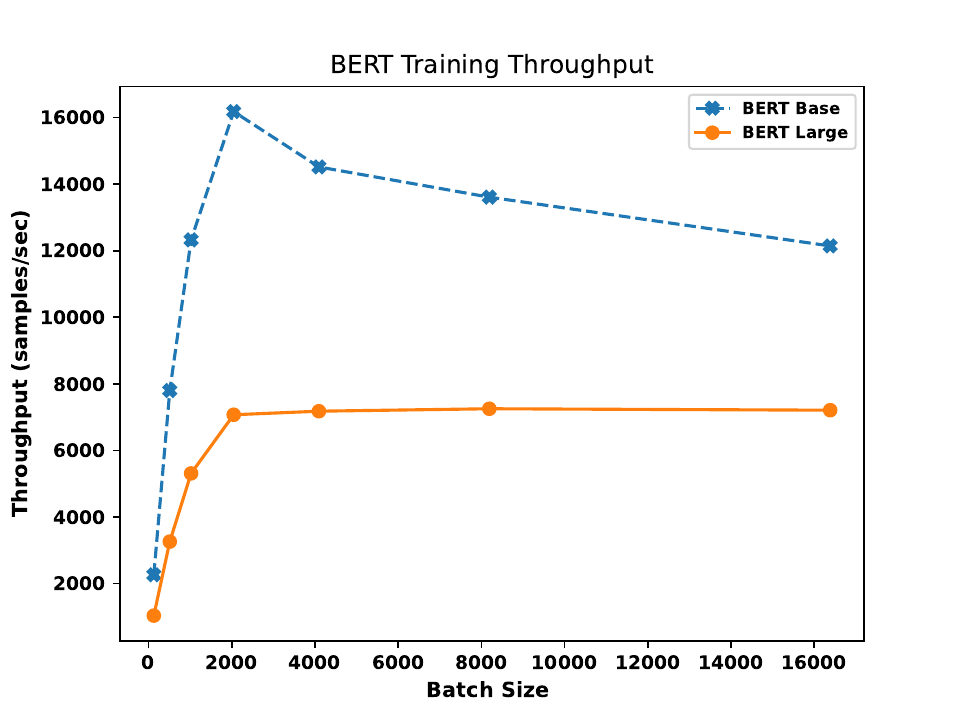}}
\caption{BERT training throughput analysis. Training throughput is measured in samples/sec. Batch sizes are all powers of 2}
\label{fig:bert_train}
\end{figure}

Training experiments of BERT models were performed on the BERT-base and BERT-large variants as shown in Table \ref{tab:bert-params}. Figure \ref{fig:bert_train} shows the training performance, measured in throughput (samples/sec), of the BERT models with varying batch sizes, where the batch size is measured in the number of samples. For both BERT-base and BERT-large, the training throughput increases initially. However, the training throughput of the BERT-base model decreases after reaching the optimal batch size, while the training throughput of the BERT-large model stays roughly the same after a sharp increase in the beginning. The optimal batch size of BERT-base is 2048 and BERT-large is 8192. 

\indent In Table \ref{bert_train_time}, we report the projected training time, in seconds, to train one epoch of the SST-2 dataset on the BERT-base and BERT-large models. We calculate the training time based only on the number of training samples, not including test and development sets, and the training throughput we observed in Figure \ref{fig:bert_train}.

\setlength{\tabcolsep}{15pt}
\renewcommand{\arraystretch}{1.5}

\begingroup{}

\begin{table}[h!]
\captionof{table}{Projected SST-2 One Epoch Training Time (secs)}\label{bert_train_time}
\centering
\begin{tabular}{c|cc}
\hline
\multirow{2}{*}{Batch Size} & \multicolumn{2}{c}{BERT}    \\ \cline{2-3}
                                  & \multicolumn{1}{c}{base}   & \multicolumn{1}{c}{large}                             \\ \hline

128                             & \multicolumn{1}{c}{29.70} & 64.73           \\

512                        & \multicolumn{1}{c}{8.63} & 20.64               \\ 

1024                          & \multicolumn{1}{c}{5.46} & 12.68                      \\ 

2048                          & \multicolumn{1}{c}{4.16} & 9.52                   \\ 
4096                            & \multicolumn{1}{c}{4.64} & 9.38 \\ 

8192                    & \multicolumn{1}{c}{4.95} & 9.29                  \\ 
16384                    & \multicolumn{1}{c}{5.55} & 9.34                  \\ 
 \hline
\end{tabular}
\end{table}
\endgroup

\subsection{GPT-3 Training Performance Analysis}
\label{subsec:bert_train}

\begin{figure}[h!]
\centering

\begin{tabular}{cc}

  \hspace{-30pt}\includegraphics[width=0.28\textwidth]{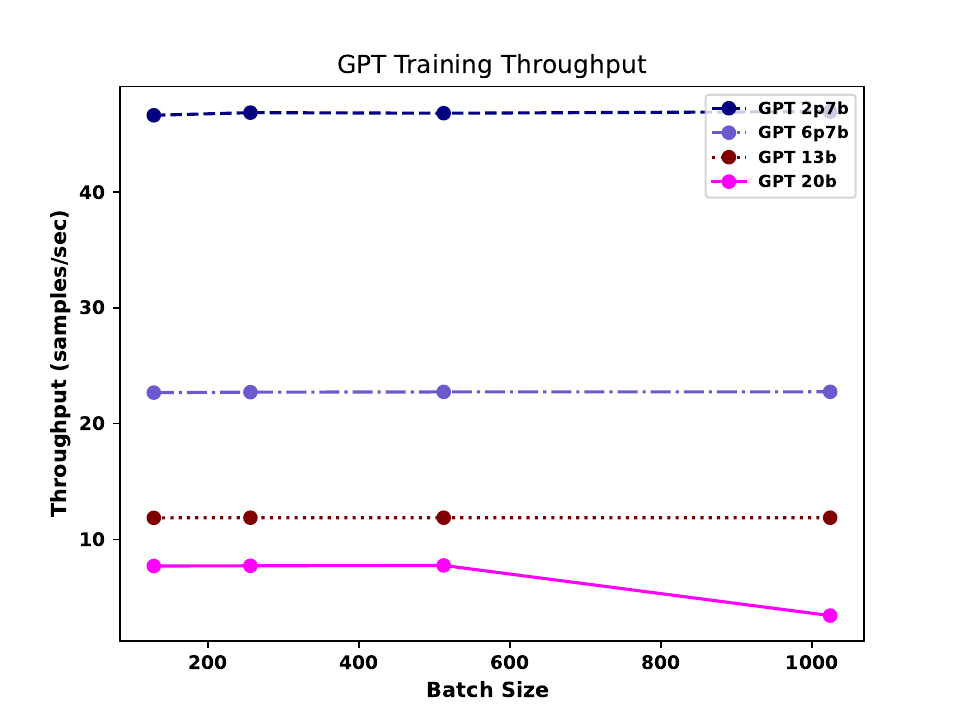}  & \hspace{-30pt}\includegraphics[width=0.28\textwidth]{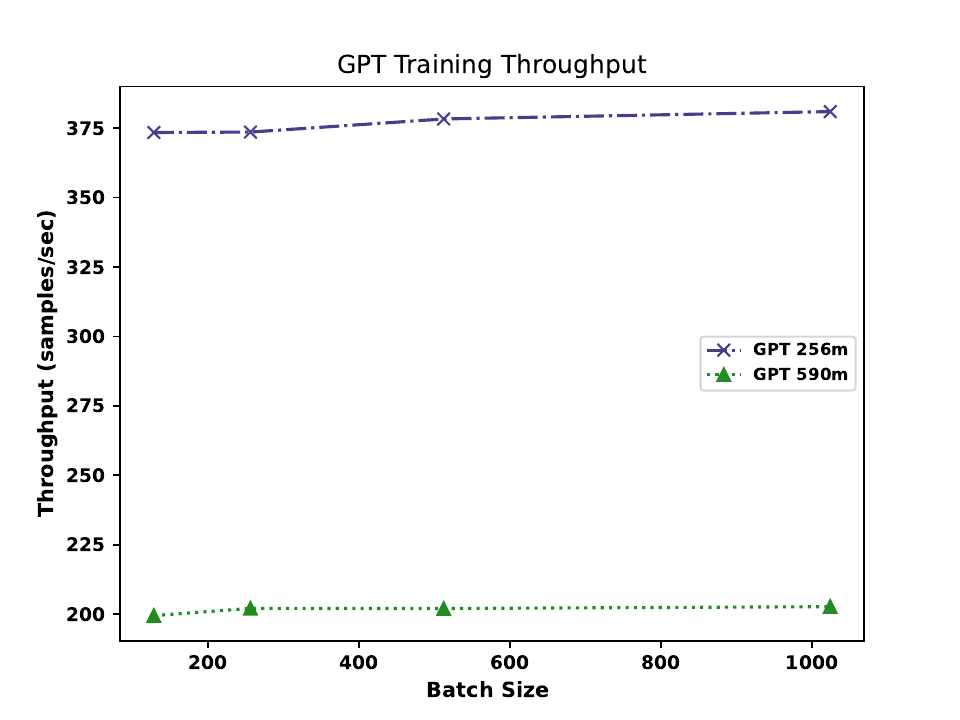}  \\
  5(a) &  5(b) \\
  [10pt]

\end{tabular}

\caption{Training throughput analysis of GPT-3 models over varying batch sizes. 5(a) shows the training throughput for GPT-3 2.7B, 6.7B, 13B, and 20B models. 5(b) shows training throughput for GPT-3 256M and 590M models. All batch sizes are measured in number of samples and are power of 2}
\label{fig:gpt_train}
\end{figure}

Training experiments of GPT-3 models were conducted on the GPT-3 model sizes indicated in Table \ref{tab:gpt-params}. Figure \ref{fig:gpt_train} shows the training throughput of each GPT-3 model size with varying batch sizes. For GPT-3 256M model, the training throughput increases as the batch size increases. Larger batch sizes can more effectively utilize the available high bandwidth memory. This facilitates more computations within each core, exploiting the large compute resources available in the WSE. For GPT-3 20B size, training throughput drops at large batch sizes. This is due to the fact that the memory bandwidth on the chip is fully utilized and data transfer dominates the compute time during training. For other GPT-3 model sizes, the training throughput stays roughly the same across model sizes. This highlights the WSE's unique capacity to scale training for large models and batch sizes without a drop in throughput. 

\indent In Table \ref{pile_train_time}, we provide the projected training time, in hours, to train one epoch of the entire PILE dataset \cite{biderman2022datasheet}. The projected training time includes one pass of the train, test, and development sets. We calculate the training time based on the training throughput we observed in Figure \ref{fig:gpt_train} and the total number of samples reported in the PILE datasheet, 211,043,181 samples. We provide the training time for batch size of 128 and 256 as the throughput is similar for other batch sizes and 128 and 256 are common choices of batch size.

\setlength{\tabcolsep}{15pt}
\renewcommand{\arraystretch}{1.5}

\begingroup{}

\begin{table}[h!]
\captionof{table}{Projected PILE One Epoch Training Time (hours)}\label{pile_train_time}
\centering
\begin{tabular}{c|cc}
\hline
\multirow{2}{*}{Model} & \multicolumn{2}{c}{Batch Size}    \\ \cline{2-3}
                                  & \multicolumn{1}{c}{128}   & \multicolumn{1}{c}{256}                             \\ \hline

256 M                             & \multicolumn{1}{c}{157} & 157           \\

590 M                        & \multicolumn{1}{c}{294} & 290               \\ 

2.7 B                          & \multicolumn{1}{c}{1258} & 1252                      \\ 

6.7 B                           & \multicolumn{1}{c}{2586} & 2581                   \\ 
13 B                            & \multicolumn{1}{c}{4943} & 4939 \\ 

20 B                    & \multicolumn{1}{c}{7613} & 7594                  \\ 
 \hline
\end{tabular}
\end{table}
\endgroup

\subsection{BERT Inference Performance Analysis}
\label{subsec:bert_inference}

\begin{figure}[htbp]
\centerline{\hspace{-20pt}\includegraphics[width = \linewidth]{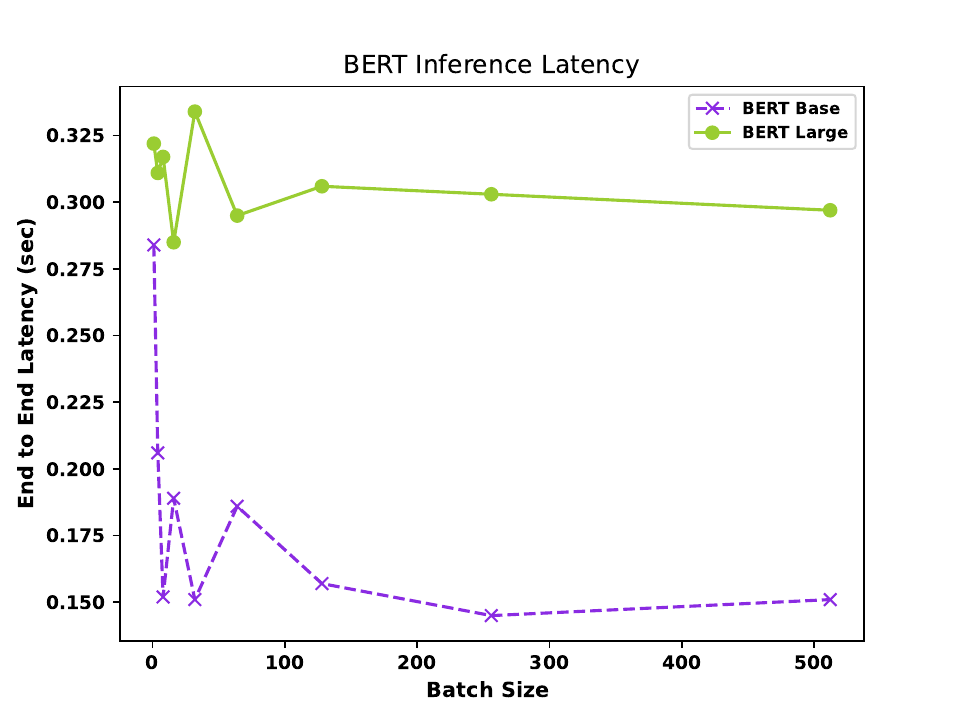}}
\caption{BERT end-to-end inference latency analysis.}
\label{fig:bert_inference}
\end{figure}

Figure \ref{fig:bert_inference} shows the inference latency for BERT models over varying batch sizes. Our experiment results indicate that the BERT-base model latency does not vary by much for all batch sizes except batch size of 1. Likewise, BERT-large model latency does not change much after a certain batch size. However, recall that end-to-end latency is measured as the duration it takes for the output of the entire batch to be produced. Therefore, this indicates that performing inference for larger batch sizes is beneficial on the Cerebras WSE platform as the end-to-end latency does not vary by much with increasing batch size. This reduces the average inference latency per sample for larger batch sizes, as is expected for high-compute high-bandwidth systems like Cerebras. 

\subsection{Roofline Models}
\label{subsec:roofline}

\begin{figure}[htbp]
\centerline{\includegraphics[width = \linewidth]{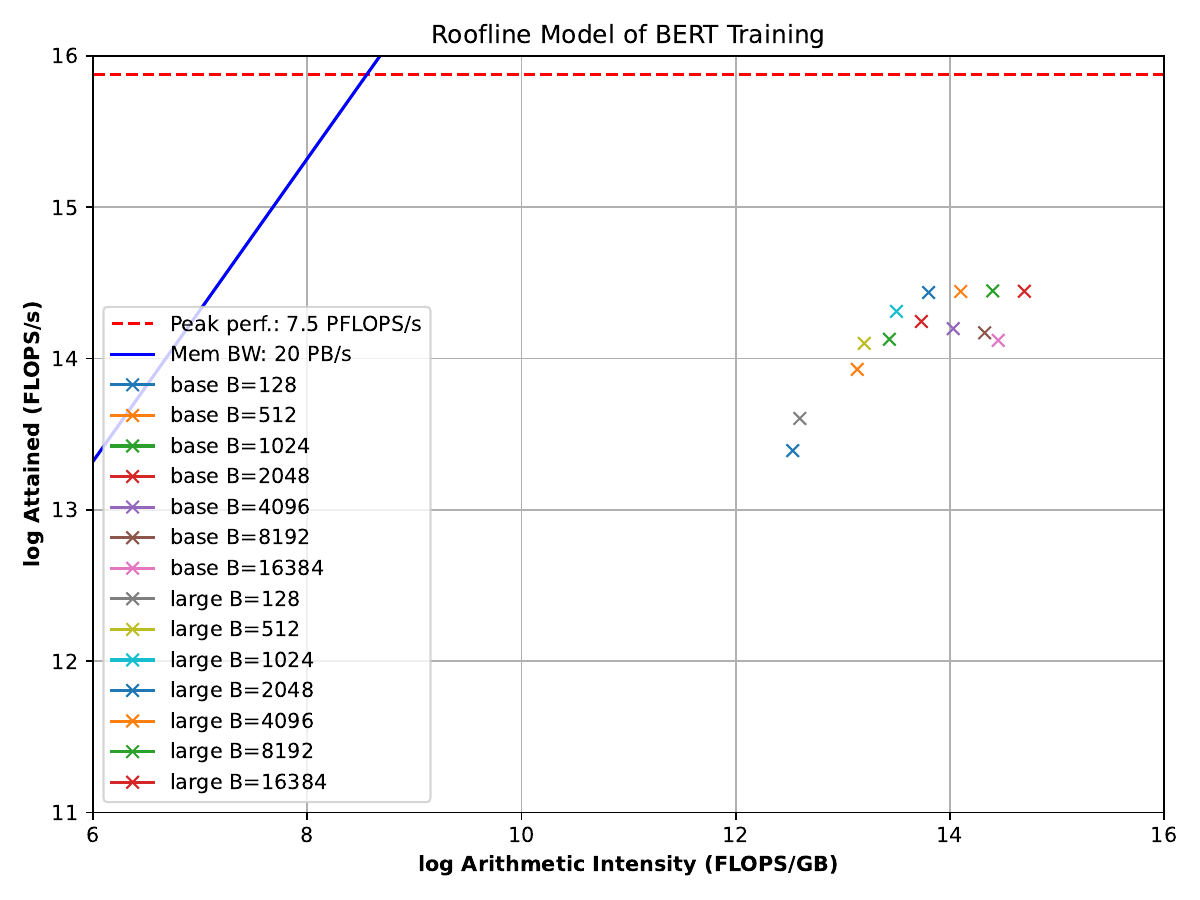}}
\caption{BERT roofline analysis. Axes are in log scale. }
\label{fig:bert_roofline}
\end{figure}

\begin{figure}[htbp]
\centerline{\includegraphics[width = \linewidth]{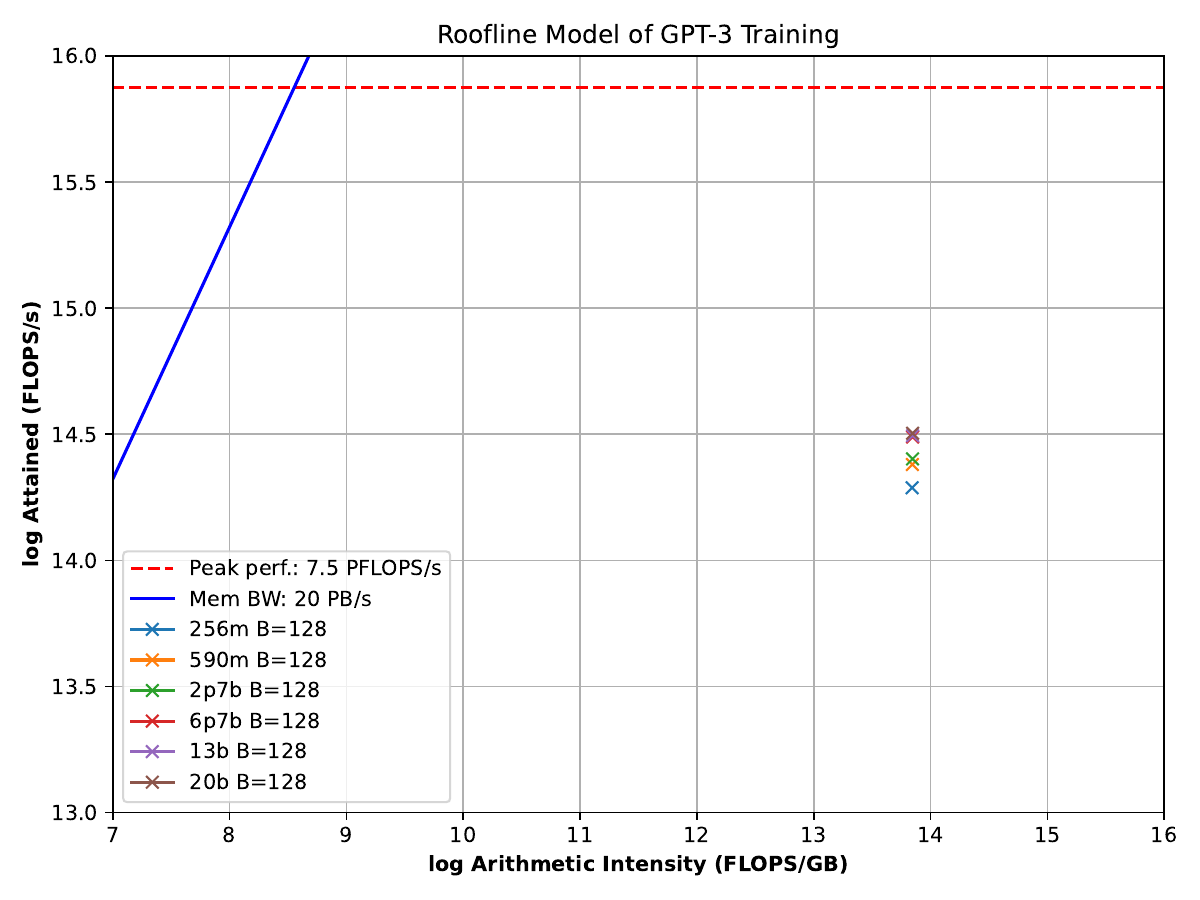}}
\caption{GPT-3 roofline analysis. Axes are in log scale. Only batch size of 128 is investigated.}
\label{fig:gpt_roofline}
\end{figure}

Based on our experimental results, we completed roofline model analysis for BERT and GPT-3 training, shown in Figure \ref{fig:bert_roofline} and Figure \ref{fig:gpt_roofline}. Our results show that both BERT-base and BERT-large training on all the batch sizes we investigated operate in the compute-bound region. Moreover, all GPT-3 training for batch size of 128 operate in the compute-bound region. This highlights that training large models on the Cerebras WSE is scalable, with Cerebras WSE scaling the memory wall for LLM training through its unique architecture. 

\section{Conclusion and Future Work}
\label{sec:conclusions}

In this paper, we examined the training throughput of BERT and GPT-3 training across commonly used batch sizes and model sizes, as well as inference latency for BERT. We also investigated the roofline model of BERT and GPT-3 training, gaining insights in the scalability and potential of the Cerebras WSE for accelerating LLM training and inference. Investigating Cerebras WSE's performance for more models, especially computer vision models, will be a promising area for exploration because of its widespread applications.

\section*{Acknowledgment}
This work was performed on a CS-2 platform at the University of Southern California supported by the US Army DURIP program. 
This work was also supported in part by DEVCOM Army Research Lab (ARL) under grant W911NF2320186. We would like to thank the USC Center for Advanced Research Computing (CARC) for their support in installing and accessing the CS-2 platform and the Cerebras staff for their continued engineering support and helpful discussions.




%

\bibliographystyle{IEEEtran}
\bibliography{ref}

\end{document}